\documentclass[twocolumn,prl]{revtex4}

\ifx\pdfoutput\undefined
  \usepackage[dvips]{graphicx}
\else
  \usepackage[pdftex]{graphicx}
\fi

\usepackage{dcolumn}
\usepackage{amsmath}
\usepackage{latexsym}
\usepackage{units}

\newcommand{\ie}{i.e.}
\newcommand{\eg}{e.g.,}
\newcommand{\etl}{{\em et al.}}

\begin{document}

\title[Short Title]{Subgap conductivity in SIN-junctions of high barrier transparency}
\author{S.~V.~Lotkhov}
\email[Electronic mail:~]{Sergey.Lotkhov@ptb.de}
\author{D.~V.~Balashov}
\altaffiliation[Permanent address: ] {Institute of Radio
Engineering and Electronics, Russian Academy of Science,
Mokhovaya 11, 101999, Moscow, Russia}
\author{M.~I.~Khabipov}
\author{F.~-I.~Buchholz}
\author{A.~B.~Zorin}
\affiliation{Physikalisch-Technische Bundesanstalt, Bundesallee
100, 38116 Braunschweig, Germany}

\date{\today}

\begin{abstract}
We investigate the current-voltage characteristics of
high-transparency superconductor-insulator-normal metal (SIN)
junctions with the specific tunnel resistance $\rho \lesssim
\unit[30]{\Omega \times \mu m^2}$. The junctions were fabricated
from different superconducting and normal conducting materials,
including Nb, Al, AuPd and Cu. The subgap leakage currents were
found to be appreciably larger than those given by the standard
tunnelling model. We explain our results using the model of
two-electron tunnelling in the coherent diffusive transport
regime. We demonstrate that even in the high-transparency
SIN-junctions, a noticeable reduction of the subgap current can be
achieved by splitting a junction into several submicron
sub-junctions. These structures can be used as nonlinear
low-noise shunts in Rapid-Single-Flux-Quantum (RSFQ) circuitry
for controlling Josephson qubits.
\end{abstract}

\maketitle

\section{Introduction}

In the past decades, the mechanisms of the conductivity through
the superconductor-insulator-normal metal (SIN) tunnel interface
have been extensively studied. Due to the superconducting energy
gap $\Delta$ in the quasiparticle density of states, the current
through such an interface is generally suppressed at low
temperatures, $k_{\text B}T \ll \Delta$, and subgap bias voltages,
$V \lesssim V_{\text g} \equiv \Delta/e$, giving rise to a
low-voltage nonlinearity in the $IV$-curve \cite{Tinkham}.
Recently this, the most common property of SIN-contacts has been
found to be rather useful, \eg{} for reducing generation of
quasiparticles by building the current shunts into the
Josephson-junction qubits \cite{Banishing} or for avoiding an
extra decoherence \cite{Zorin-SIN} in all-Josephson RSFQ-qubit
integrated systems \cite{Semen}.

In particular, the thermal noise produced by the resistively
shunted Josephson junctions in a standard RSFQ-circuitry, brings
into being an additional source of qubit decoherence
\cite{Zorin-SIN}. It was shown however that this effect can be
minimized, by using the nonlinear damping that is provided by an
SIN-junction of low asymptotic resistance, $R_{\text N}$, and a
high value of the nonlinearity parameter $\eta \equiv \left[ G(0)
R_{\text N} \right]^{-1}$, where $G(0)$ is the zero-bias
conductivity of the shunt. The dynamics of SIN-shunted Josephson
junctions has recently been analyzed in detail on the basis of a
standard tunnelling model of the SIN-junction (see
Ref.~\cite{Zorin-microscopic}). The feasibility was demonstrated
to ensure both a high damping at the characteristic Josephson
frequency $\omega_{\text C}~\equiv~ 2eV_{\text C}/\hbar$, where
$V_{\text C}~=~I_{\text C}R_{\text N}$ is the characteristic
voltage and $I_{\text C}$ the critical current, and, due to the
high impedance of the SIN-junction at subgap voltages, a
sufficiently low noise at qubit frequencies of typically up to
$\sim\unit[10-30]{GHz} \ll \Delta/h $.

The first experiments on SIN-shunted Nb SIS Josephson junctions
\cite{Zorin-SIN} have proven the feasibility of the overdamped
regime at $T = \unit[4.2]{K}$. At the same time, however, an
unexpectedly weak subgap nonlinearity of the SIN-junctions was
observed. In particular, the value of $\eta \approx 6.2$, measured
at $T = \unit[1.4]{K}$ in an SIN-junction with a specific tunnel
resistance of $\rho \approx \unit[135]{\Omega \times \mu m^2}$,
was much smaller than the values $\eta \gtrsim 100$ which are
typical of opaque barriers (see, \eg{}
Refs.~\cite{Opaque,PothierGueron}), and also far below the simple
estimates made for the tunnelling model, yielding a large subgap
suppression factor $\propto {\rm exp}(-\Delta/k_{\text B}T)$
\cite{Tinkham}. In view of the previous data, it would be an even
more challenging task to achieve a high nonlinearity in
SIN-junctions of very low specific resistance, \eg{} $\rho
\lesssim \unit[30]{\Omega \times \mu m^2}$ for the SIN-junctions
based on superconducting Al \cite{Zorin-SIN}, as required for
RSFQ operation at qubit temperatures $T \lesssim \unit[50]{mK}$.

\begin{figure}[t]
\begin{center}
\leavevmode
\includegraphics[width=3.2in]{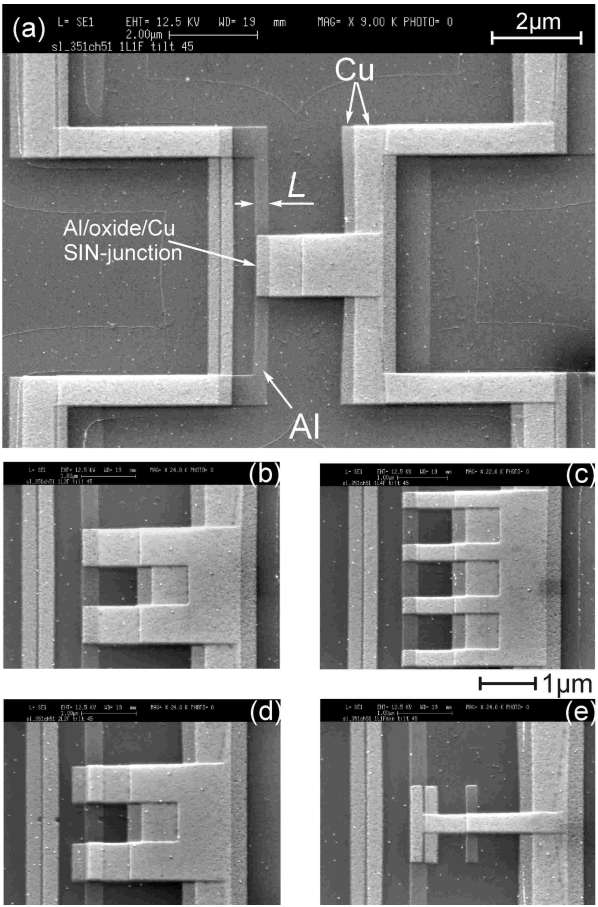}
\caption {SEM-micrographs of the sample 1L1F (a) and blow-up of
the junctions 1L2F (b), 1L4F (c), 2L2F (d), and 1L1Fmin (e), of
different geometries. The total tunneling area is the same for
each sample, producing similar asymptotic resistances (except for
the "2Lx"-samples [(d) in this Figure], as measured in a 4-point
configuration.} \label{Fig1}
\end{center}
\end{figure}
In this paper, we address subgap conductance phenomena in
SIN-junctions of very high transparency. In particular we found
that the degree of nonlinearity of the measured $IV$-curves is
consistent with predictions of the two-electron tunnelling model
which has been developed in a ballistic regime by Blonder,
Tinkham and Klapwijk (BTK) \cite{BTK} and modified by Hekking and
Nazarov \cite{HekNazPRL,HekNazPRB} to account also for
interference contributions due to coherent electron diffusion. In
earlier experiments, the spatial coherence was clearly
demonstrated by Pothier \etl{} \cite{PothierGueron} in the form of
a periodical dependence of the low-bias conductance of the
so-called NS-QUID interferometer on the magnetic flux. A
spatially coherent enhancement of the two-electron tunnelling
amplitude makes the junction conductance topology-dependent. This
effect was observed as {\em zero-bias conductance anomaly\/} in a
thin-film planar NS-interface, which was compared in
Ref.~\cite{Xiong} with the BTK ballistic transport through an
edge-type NS-contact. In our experiment, we compare the values of
the nonlinearity parameter $\eta$ for different junction
topologies, varying the junction sizes and/or the width of the
adjacent diffusive electrode. We show that a sufficiently strong
nonlinearity can be obtained in submicron SIN-contacts with even
high-transparency barriers, \ie{} in the case of interest for
RSFQ-qubit applications.

\section{Design of samples}

Two techniques were applied for fabricating the samples. The large
reference junctions "NbAuPd" and "NbAl" (see Table~\ref{Sampa})
were made in a standard sandwich technique which is usually
applied for RSFQ-devices (see, \eg{}
Refs.~\cite{Dolata,Balashov}), on the basis of a Nb trilayer,
patterned with the help of optical lithography. Both samples had
a sophisticated multilayered structure which made a particular
analysis of the SIN-junction component difficult. For example,
the junction "NbAl" consisted of the metal stack with a
pronounced proximity effect between the AuPd and the top Nb
layers, and, possibly, between the bottom Nb and the covering Al,
separated by a thin oxide barrier only. These Nb-based samples
were characterized in a He$^4$-cryostat at temperatures down to
$\unit[1.4]{K}$. The data for these two samples, shown in
Table~\ref{Sampa}, represent our rough estimations.

For the sake of simplicity and flexibility of the design, all
other junctions were fabricated by means of Dolan's shadow
evaporation process \cite{Dolan}. The complete structures were
deposited {\em in-situ\/} from the e-gun through the single
e-beam-patterned PMMA/Ge/Copolymer mask with free-hanging
bridges. The junctions were formed on the overlapping areas  of
the first and the subsequent shifted layers, deposited through
the same openings in the mask but at different tilting angles of
the wafer in respect to the incident material flow. The
$IV$-curves were measured at $T \lesssim \unit[0.05]{K} $ in a
dilution refrigerator.

\begin{figure}[b]
\begin{center}
\leavevmode
\includegraphics[width=3.2in]{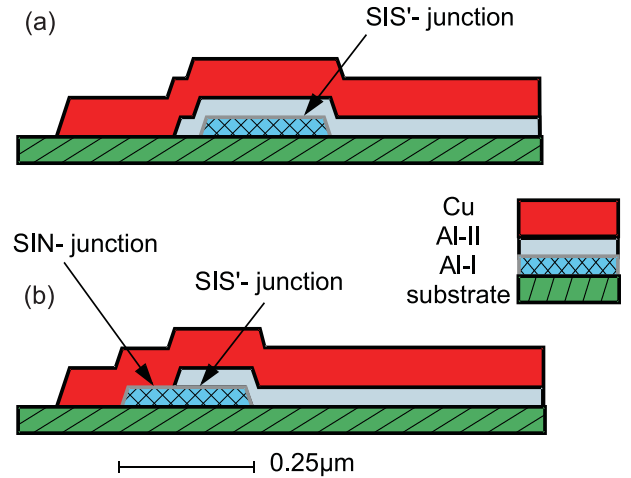}
\caption {Cross-section sketch of the samples SISN (a) and
SISNSIN (b). In the sample SISNSIN, the junction was a
composition of an SIN and an SIS' sub-junction, both having the
same Al-I bottom layer (see text). } \label{SISN}
\end{center}
\end{figure}

The bottom layer of all shadow-evaporated junctions was made of
aluminum, with a thickness of $d_{\rm B} = \unit[30]{nm}$, and
oxidized in a very mild regime (see below) to form a thin tunnel
barrier. To be able to put several different structures on the
same chip, we used besides a two-layer also a three-layer design,
with Cu-, Cu/Cu- or Al/Cu-layers building the top electrodes of
the junctions. The junctions "SINTOP" and "SINBOT" were simple
two-layer structures with the tunnel areas differing by the
factor of two, both oxidized for 5 min at an oxygen pressure of
$P_{\rm O_2} =\unit[1]{Pa}$, which resulted in a specific barrier
resistance $\rho = \unit[27]{\Omega \times \mu m^2}$. A somewhat
more complicated layout was used for the junctions "1L1F" to
"2L4F" which were fabricated on the next wafer. The top junction
electrode was made either of a single layer [{\em the top\/} Cu
film in the "1Lxx"- junctions, see, \eg{}
Fig.~\ref{Fig1}(a-c),(e)] or of a double-layer composition [a
composition of {\em the middle and the top\/} Cu films in the
"2Lxx"-junctions, see, \eg{} Fig.~\ref{Fig1}(d)], evaporated in
two consecutive steps, with an additional tilting of the wafer in
between. Accordingly, two different thicknesses of the top
electrodes, $d_{\rm T} = \unit[30]{nm}$ and $\unit[60]{nm}$, were
obtained. This wafer was oxidized for 5 min at lower pressure,
$P_{\rm O_2} =\unit[0.1]{Pa}$, in order to obtain an even more
transparent barrier. Although all junctions had the same Al
bottom layer with the same oxidation step, two different values
of $\rho$ were found: $\rho = \unit[15]{\Omega \times \mu m^2}$
for the "1Lxx"-junctions, and $\rho = \unit[5]{\Omega \times \mu
m^2}$ for the "2Lxx"-junctions. A similar effect was observed for
the samples "SISN" and "SISNSIN". In our opinion, this variation
is specific to low-dose oxidation processes and arises due to a
prolonged oxidation of the Al by adsorbed oxygen in the areas
which are not being covered by the middle layer of Cu.

The junctions "1L2F" to "2L4F", "SISN" and "SISNSIN" were
designed in a "split-finger" layout [see Figs.~\ref{Fig1}(b-d)]
by subdividing the junction into $N = $~2 or 4 narrower
sub-junctions with a partial width $W$ each, as listed in
Table~\ref{Sampa}. The neighboring fingers of Cu formed, together
with the superconducting Al sections, the closed loops of the area
$\sim 1 \times \unit[1]{\mu m^2}$. In contrast to the other
samples, the normal metal electrodes in the junctions "1L1Fmin"
and "1L2Fmin" were made in a special "electron confinement" shape
[Fig.~\ref{Fig1}(e)]: The width of the adjacent wire was reduced
further to 1/4 of $W$ while the tunnel area was kept identical to
that in the samples "1L1F" and "1L2F", respectively.

The top electrode of junctions in the samples "SISN" and
"SISNSIN", both fabricated on the same wafer, consisted of two
different metals, Al and Cu, which formed slightly shifted shadows
in such a way that, in the sample "SISN", the junction was
designed to be effectively of type
"superconductor-insulator-superconductor" with a proximity-reduced
gap (SIS'). In the sample "SISNSIN", the junction of the same
total area was made up of such an SIS' to the half with an
Al/oxide/Cu SIN-contact (see the cross-sections in
Fig.~\ref{SISN}).

\begingroup
\squeezetable
\begin{table*}[t]
\caption{\label{Sampa} Parameters of the junctions}
\begin{ruledtabular}
\begin{tabular}{p{1.4cm} @{\vline} c p{2.8cm} p{1.9cm} c c c c c}
Sample code & Materials & Tunnel junction area, $N$$\times$
($L$$\times$$W$) ($\mu m^2$) & Thicknesses, $d_{\text B}$,
$d_{\text T} ...$ (nm) & $T$ (K)& $R_{\text N}
(\Omega)$ & $\rho$$(\Omega$$\times$$\mu m^2)$ & $\left[ G(0) \right]^{-1}(\Omega)$ & $\eta$ \\
 \hline
  &&&&&&&\\
  NbAuPd  & Nb/AlO$_{\text x}$/AuPd & 24                              & 180,100 & 1.4  & 16 & 380 & 95    & 6\\

  SINTOP  & Al/AlO$_{\text x}$/Cu   & 0.5$\times$2                  & 30,50   & 0.03 & 27 & 27  & 140   & 5.1\\
  SINBOT  & Al/AlO$_{\text x}$/Cu   & 0.5$\times$1                  & 30,50   & 0.03 & 54 & 27  & 360   & 6.6\\

  1L1F    & Al/AlO$_{\text x}$/Cu   & 0.24$\times$2                 & 30,30   & 0.05 & 29 & 15 &   90 & 3.1 \\
  1L1Fmin\footnotemark[1] & Al/AlO$_{\text x}$/Cu   & 0.24$\times$2                 & 30, 30   & 0.05 & 32 & 15 &   64 &   2 \\
  1L2F    & Al/AlO$_{\text x}$/Cu   & 2$\times$(0.24$\times$1)    & 30,30   & 0.05 & 28 & 15 &   88 & 3.1 \\
  1L2Fmin\footnotemark[1] & Al/AlO$_{\text x}$/Cu   & 2$\times$(0.24$\times$1)    & 30,30   & 0.05 & 42 & 15 &  113 & 2.6 \\
  1L4F    & Al/AlO$_{\text x}$/Cu   & 4$\times$(0.24$\times$0.5)  & 30,30   & 0.05 & 31 & 15 &  177 & 5.7 \\
  2L2F    & Al/AlO$_{\text x}$/Cu   & 2$\times$(0.24$\times$1)    & 30,60   & 0.05 & 11 &  5 &   22 &   2 \\
  2L4F    & Al/AlO$_{\text x}$/Cu   & 4$\times$(0.24$\times$0.5)  & 30,60   & 0.05 & 11 &  5 &   37 & 3.4 \\

  SISN   & Al/AlO$_{\text x}$/Al/Cu& 2$\times$(0.25$\times$0.5)   & 30,30,60 & 0.026 & 44  & 11   & 1.9k & 43 \\
  SISNSIN& Al/AlO$_{\text x}$/Al/Cu+Al/AlO$_{\text x}$/Cu & 2$\times$(0.12/0.12\footnotemark[2]$\times$0.5) & 30,30,60 & 0.026 & 63  & 11/22\footnotemark[2] & 3.3k & 52 \\

  NbAl   & Nb/AlO$_{\text x}$/Al/AlO$_{\text x}$/AuPd/Nb   & 100      & 90,100,100,500  & 1.45 & 1.4 & 150 & 1.4k & $10^3$\\

\end{tabular}
\end{ruledtabular}
\footnotetext[1]{In these samples, the normal electrode is of
"electron-confinement" shape (see the text).} \footnotetext[2]{For
SISN- and SIN-partial junctions, respectively.}
\end{table*}
\endgroup

\section{Results and discussion}

\begin{figure}[t]
\begin{center}
\leavevmode
\includegraphics[width=3.2in]{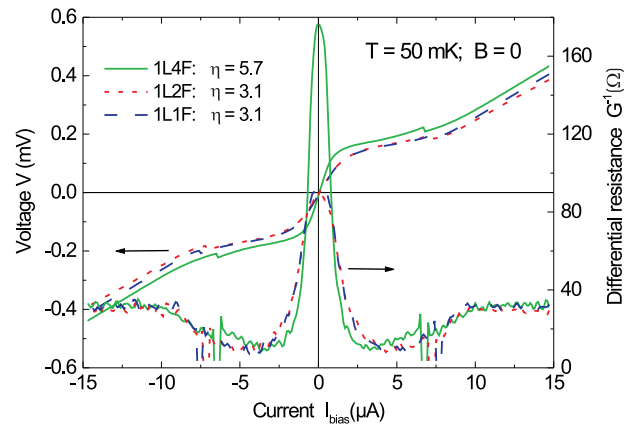}
\caption {Comparative $IVC$-plot for the samples "1L1F", "1L2F",
and "1L4F". Whereas the asymptotic slopes of all three samples are
similar, the zero-bias resistance of the sample "1L4F", which is
the one with the narrowest wires and junctions, exceeds $R_{\rm
J}$ of the other samples noticeably, indicating the geometric
effect for the two-electron conductance. The small kicks on both,
positive and negative branches of the $IV$-curves around $V = \pm
\Delta/e$ arise due to self-heating effects in the junctions
(see, \eg{} Ref.~ \cite{Overheating}).} \label{ivc1L1F-4F}
\end{center}
\end{figure}

Typical $IV$-characteristics of high-transparency SIN-junctions
are shown in Fig.~\ref{ivc1L1F-4F}. As expected, significant
subgap leakage currents were observed in most of the
SIN-junctions, even at the temperature $T = \unit[50]{mK} \ll
\Delta/k_{\text B}$, resulting in a strong smearing of the
superconducting gap corner and in rather moderate values of $\eta
\sim $1--10 (see Table~\ref{Sampa}). The appreciable excess
subgap current at low temperatures is commonly attributed to the
resonant two-electron transport channel due to Andreev reflection
\cite{Andreev}. To interpret our data, we first make estimations
within the simplified ballistic BTK model, which put aside the
electron scattering effects in the electrodes. Using Eq.(17) of
\cite{BTK}, it is straightforward to derive simple expressions
for the subgap conductance at $T=0$:
\begin{equation}
\label{BTK-conductivity} R_{\rm N} G(V) = \frac{{2\Delta ^2 \left(
{1 + Z^2 } \right)}}{{(eV)^2 + \left[ {\Delta ^2  - (eV)^2 }
\right]\left[ {1 + 2Z^2 } \right]^2 }}
\end{equation}
and the BTK nonlinearity parameter:
\begin{equation}
\label{BTK-ratio} \eta_{\rm BTK} \equiv \left[ R_{\rm N} G(0)
\right]^{-1} = \frac{{\left( {1 + 2Z^2 } \right)^2 }}{{2(1 + Z^2
)}},
\end{equation}
whereby the dimensionless barrier strength $Z$ \cite{BTK}
characterizes the transparency of the NS interface, $Z = k_{\rm
F}H/2\epsilon_{\rm F}$, $\epsilon_{\rm F}$ and $k_{\rm F}$ are the
Fermi energy and wavenumber, respectively and $H$ enters the
phenomenological repulsive barrier potential $H \delta(x)$. For a
thin tunnel barrier of the simplest, rectangular shape with a
height $U > \epsilon_{\rm F}$ and a thickness $a \sim 2\pi k_{\rm
F}^{-1} = \lambda_{\rm F}$, which is the Fermi wavelength, $Z$
can be expressed as
\begin{equation}
\label{Ztunnel}  Z = \frac{k_{\rm F} U}{2\epsilon _{\rm F}}\left[
\frac{{\rm sinh}~(a \kappa) }{\kappa } \right], ~  \kappa  =
k_{\rm F} \sqrt {\frac{U - \epsilon _{\rm F} }{\epsilon _{\rm F}}}
\end{equation}
by matching the well-known plane-wave transmission coefficient
(see, \eg{} \cite{LandauLifshits}) with that of BTK, $C =
(1+Z^2)^{-1}$, which is provided in Ref.~\cite{BTK} for electrons
in the normal state.
The straightforward evaluation using Eqs.~(\ref{BTK-ratio}) and
(\ref{Ztunnel}) and the parameters of the idealized Al/AlO$_{\rm
x}$-barrier, $U-\epsilon_{\rm F} \sim \unit[2]{eV}$ (see, \eg{}
Refs.~\cite{McBride,Maezawa}) and $a_{\rm 0} \approx
\unit[5]{\AA}$ (see the discussion below), as well as
$\epsilon_{\rm F} = \unit[11.4]{eV}$ and $k_{\rm F} = \unit[1.75
\times 10^{10}]{m^{-1}}$, which are typical for Al, results in the
values $Z \approx 27$ and $\eta_{\rm BTK} \approx 1500$ and
predicts the suppression of the subgap conductivity much stronger
than found in our experiment.

Important corrections can be made to the calculations if we take
a note of the fact that the barriers under consideration are as
thin as a single monolayer of Al oxide. The value of $a_{\rm 0}$
accepted above presents an average barrier thickness of the plane
junction with the specific capacitance $c \approx
\unit[75]{fF/\mu m^2}$ (derived from the Coulomb offset voltage
of the similarly oxidized single-electron transistors
\cite{LotkhovJAP2004}, cf. also \cite{Zant}), and the dielectric
constant $\varepsilon_{\rm r} \approx 4$ \cite{Maezawa}. For such
thin oxide layers, it is plausible to expect considerable local
deviations of the relative barrier thickness, even for the
cleanest tunnel interface.

Possible models for the fine structure of very thin barriers have
recently been discussed for junctions of type Nb/AlO$_{\rm x}$/Nb
with current densities in a wide range, $J_{\rm c} \sim
\unit[0.1-20]{kA/cm^2}$ \cite{Maezawa,Zant,Kleinsasser}. In
Refs.~\cite{Zant,Kleinsasser}, the barrier is supposed to consist
of different areas with a certain number of oxide monolayers,
including a statistically large amount of local pinholes of
lateral size $\sim \unit[1]{nm}$, with no single monolayer of
oxide at all, dominating in the current of the whole junction. In
contrast to this, the model of a uniform, continuously growing
barrier was suggested by Maezawa \etl{} \cite{Maezawa} as a more
realistic, predicating upon that oxide is formed on top of a
granular Al film with a locally different surface orientation of
the crystal planes. Our argument is that small, but finite
values, $Z > 1$, corresponding to our experimental value $\eta
\gtrsim 2$ (see Fig.~6 of Ref.~\cite{BTK}), rather relate to the
picture of a thin, but finite tunnel barrier, at least on the
mesoscopic scale $\sim L,W$. Moreover, on the microscopic scale
$\sim \lambda_{\rm F}$, the ballistic transport model accounting
for the pinhole effects is not adequate; the relevant approach
here relates to the electron diffraction, which makes the
tunnelling picture non-local \cite{Diffraction}.

To build a quantitative model, we assume a uniform distribution
of the barrier thickness around the mean value $a_{\rm 0}$, \ie{}
$a_{\rm 0}(1-\delta ) < a < a_{\rm 0}(1+ \delta )$, with a
constant probability density $\omega \equiv 1/2 \delta $, and
consider an independent electron transport through the areas
within the junction with different local barrier strengths. The
effective zero-bias nonlinearity can then be found as the ratio
$\eta_{\rm av}$ of the average asymptotic $\left\langle R_N^{-1}
\right\rangle$ and the zero-bias conductivities $\left\langle G(0)
\right\rangle$:
\begin{equation}
\label{BTK-average} \eta_{\rm av} \equiv \frac{\left\langle
R_{\rm N}^{-1} \right\rangle}{\left\langle G(0)\right\rangle} =
\frac{1}{2}\frac{{\left\langle {(1 + Z^2 )^{ - 1} } \right\rangle
}}{{\left\langle {\left( {1 + 2Z^2 } \right)^{ - 2} }
\right\rangle }},
\end{equation}
best matching the experimental range of $\eta \sim $~2--7 in
Al-based SIN junctions for $\delta \sim $~0.8--0.9, \ie , for the
realistic barriers with a thickness of up to two oxide monolayers.

So far we have been able to obtain a reasonable, though rough,
estimation of the zero-bias nonlinearity of SIN junctions within
the scope of the ballistic BTK model. A more detailed insight can
be gained, by comparing the subgap transport properties in SIN
junctions of the same transparency but of different size. For
example, in Fig.~\ref{ivc1L1F-4F} the voltage-current
characteristics of the single-junction structure "1L1F"
[Fig.~\ref{Fig1}(a)] is compared to those of the split-finger
interferometer structures "1L2F" and "1L4F"
[Fig.~\ref{Fig1}(b,c)]. All three structures have the same total
tunnel area and, therefore, similar asymptotic resistances, but
they differ with regard to the number $N$ and the width $W$ of the
partial junctions. In none of the split-finger interferometer
structures, a periodic dependency of the $IV$-curves on the
magnetic field was observed, which indicates the absence of
inter-junction coherence effects such as those reported earlier
for different structure dimensions by Pothier \etl{}
\cite{PothierGueron}. This property of our structures is obviously
due to the normal-metal sections of each loop being, in total,
longer than the interference length in a diffusive conductor
$L_{\rm T} \sim \unit[1]{\mu m}$ \cite{HekNazPRL} and, when used
in practice, it makes damping on the base of the split-finger
SIN-junction flux insensitive. Independent tunnelling in the
sub-junctions enables a parallelization of the SIN-contacts, when
a low shunting resistance is required.

For the larger junctions "1L1F" and "1L2F", with $W \gtrsim
\unit[1]{\mu m}$, the value of $\eta = 3.1$ does not vary with
the junction width. On the contrary, for the parallel connection
of the four smaller, $\unit[0.5]{\mu m}$-wide sub-junctions in
sample "1L4F", the value of $\eta = 5.7$ is nearly twice as
large. The same tendency is to be seen for the samples "2L2F" and
"2L4F" with slightly different parameters (see Table~\ref{Sampa}).
The observed dimensional effect can be interpreted following the
model of Hekking and Nazarov \cite{HekNazPRL,HekNazPRB} which
accounts for spatially coherent contributions to the Andreev
conductance due to two-electron diffusion. According to this
model, the coherence factors, and thus the subgap conductance, are
expected to increase with the junction size $W$ until saturation
occurred at $W \sim L_{\rm T}$. For small junctions of size $W <
L_{\rm T}$, this effect can be evaluated using the relation
Eq.~(12) of Ref.~\cite{HekNazPRB} which was derived for a normal
wire in a tunnel contact to a superconducting sheet, rewritten in
the form:
\begin{equation}
\label{1D} \eta_{\rm coh} \equiv \left[ R_{\rm N} G(0) \right]
^{-1} \propto \frac {\sigma_{\rm D} d_{\rm T} \rho}{W\times L}~,
\end{equation}
where $\sigma_{\rm D}$ is the specific conductivity of the normal
electrode. In our experiment, the validity of Eq.~(\ref{1D}) was
confirmed by a simple comparison of the nonlinearity parameter
values in the junctions "1L2F" ($\eta = 3.1$), and "2L2F" with
$\eta = 2$ [the layout shown in Fig.~\ref{Fig1}(d)] with a single
and a double layer of Cu in the top electrodes, correspondingly.
The $\sim 1.5$ times larger value of $\eta$ for "1L2F" arises as
the ratio 3 of the specific barrier resistances, multiplied by the
ratio 1/2 of the thicknesses of the top electrodes.

\begin{figure}[t]
\begin{center}
\leavevmode
\includegraphics[width=3.2in]{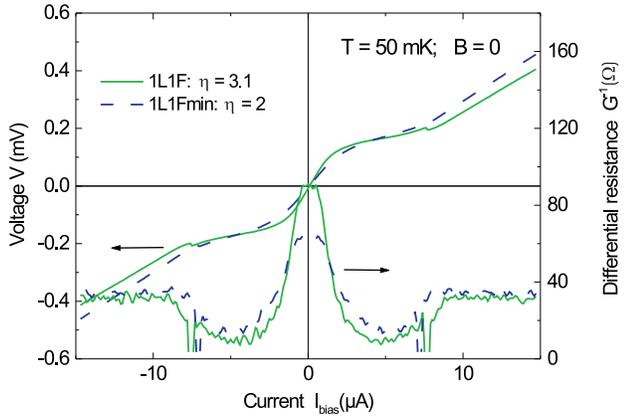}
\caption {Current-voltage  and differential resistance plots,
comparative for the samples "1L1F", with the full-width normal
wire, and "1L1Fmin" with the narrowed normal wire ("confinement"
shape).} \label{ivc1L1F-1Fmin}
\end{center}
\end{figure}

Our data also clearly demonstrate the effect of the "confinement"
shape for the normal wire. As seen in Fig.~\ref{ivc1L1F-1Fmin},
the zero bias differential resistance of such a sample "1L1Fmin"
[shown in Fig.~\ref{Fig1}(e)] is $\sim 1.5$ times lower than that
of the "plain" junction "1L1F", even despite its higher value of
$R_{\text N}$. This tendency also shows in the samples "1L2F" and
"1L2Fmin". From this, we can conclude that those configurations of
the top electrode where the diffusion volume for the
back-scattered electrons is restricted to the closer vicinity at
the tunnel barrier result in a larger Andreev conductance, which
is consistent with the predictions of the model
Ref.~\cite{HekNazPRL,HekNazPRB}.

\begin{figure}[t]
\begin{center}
\leavevmode
\includegraphics[width=3.2in]{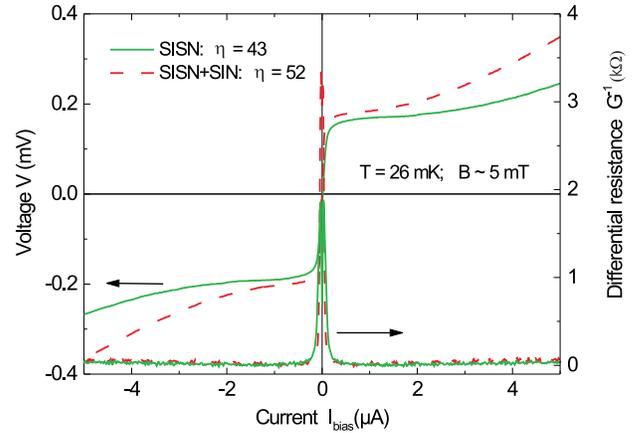}
\caption {Current-voltage  and differential resistance plots for
the samples "SISN"  and "SISNSIN". A small magnetic field, applied
perpendicularly to the sample, suppresses the critical current of
the junctions.} \label{ivc-SISN}
\end{center}
\end{figure}

A considerably higher nonlinearity of the $IV$-curves, shown in
Fig.~\ref{ivc-SISN}, was observed in the junctions "SISN" and
"SISNSIN", where the top electrode was made of a smaller-gap
superconductor, $\Delta ' \ll \Delta$, consisting of
superconducting Al, covered by a thicker layer of the normal
conductor (Cu), see Fig.~\ref{SISN}. Due to the second small gap
in the density of states, the two-electron conductance is
suppressed at small voltages $eV \lesssim \Delta '$ (see, \eg{}
\cite{Schrieffer}), while the multiple Andreev reflections are
expected to decrease at small bias as the processes of the higher
order in tunnel transparency \cite{Kleinsasser}. In experiment, a
strong current reduction was found at voltages below $\sim \Delta
/e$ (cf. also \cite{Taylor}), resulting in larger values of $\eta
= 43$ and 52 in the samples "SISN" and "SISNSIN", respectively. A
higher ratio in the latter sample can be explained by the lower
barrier transparency in the SIN partial junction, see
Eq.~\ref{1D}. A small critical supercurrent $I_{\rm c} \lesssim
\unit[300]{nA}$, otherwise observed, could be easily suppressed
below the level of fluctuations by a applying perpendicular
magnetic field of $B \sim \unit[5]{mT}$.

A much larger value $\eta = 10^3$ was measured in the sample
"NbAl", at the temperature $T = \unit[1.45]{K}$ which slightly
exceeds the superconducting transition point for Al. The layer of
Al was originally considered in this structure to form a
normal-metal electrode whereas the "S" electrode was a proximity
superconductor built of an AuPd/Nb bilayer. However, much higher
subgap nonlinearity than in the other SIN-junctions indicates, in
our opinion, that the Al electrode has a spectrum of states which
differs from that of a normal metal. For example, a small gap
could be induced in Al by the bottom film of Nb through a very
transparent separating tunnel barrier, making the junction
effectively an SIS' type. We believe that the effect of nonlinear
damping in SIS structures deserves a separate investigation.

\section{Conclusions}

We investigated the nonlinearity of the $IV$-characteristics of
SIN-junctions of high transparency, fabricated with different
topologies and composed of various materials. Our data are
consistent with the predictions of the two-electron tunnelling
model, accounting for the interference contributions in the
diffusive scattering regime. A clear enhancement of the
nonlinearity was observed in the submicron junctions with planar
dimensions smaller than the length of interference $L_{\rm T}
\sim \unit[1]{\mu m}$. We show that in a split-finger junction
layout, tunnelling in partial junctions is independent and
flux-insensitive. This property allows a parallelization of these
nonlinear and, therefore, low-noise submicron SIN-junctions. A
significant increase in the subgap resistance - compared to the
SIN-junctions of the similar transparency - was observed in
proximity-type SIS'-junctions. This could be the subject of a
special study.

The values obtained for the zero-bias and the asymptotic junction
resistances allow the SIN-junctions to be implemented in
RSFQ-Qubit circuits as low-noise shunts. For example, our
junctions with, typically, $R_{\text N} \sim
\unit[10-30]{\Omega}$ could provide a sufficient damping for
submicron Josephson junctions with $I_{\text C} \sim
\unit[10]{\mu A}$. On the other hand, sufficiently long
decoherence times, \eg{} $\gtrsim \unit[10]{\mu s}$, as estimated
in Ref.~\cite{Chiao} for a tunable flux qubit, coupled to the RSFQ
control circuit, can be achieved due to the high zero-bias
resistances, $\left[ G(0) \right]^{-1} \sim
\unit[50-100]{\Omega}$, of most of the measured SIN-junctions.

\section*{ACKNOWLEDGMENTS}
Helpful discussions with Yu.~V.~Nazarov, J.~Niemeyer,
J.~P.~Pekola, H.~Pothier and D.~Vion are gratefully acknowledged.
The work was partially supported by the EU through the projects
RSFQubit and EuroSQIP, and by Deutsche Forschungsgemeinschaft
with the project NI 253/7-1.

\end{document}